\documentclass[12pt,preprint]{aastex}

\newcommand\msun{M_{\odot}}
\newcommand\lsun{L_{\odot}}
\newcommand\rsun{R_{\odot}}
\newcommand\msunyr{M_{\odot}\,\rm yr^{-1}}
\newcommand\be{\begin{equation}}
\newcommand\en{\end{equation}}

\newcommand\mdot{\dot{M}}

\def\micron{$\mu$m }

\begin{document}

\shortauthors{Muzerolle et al.}
\shorttitle{24 \micron Detections of Young Low-mass Objects}

%\title{24 \micron Detections of Disks around Very Low-mass Stars and Brown Dwarfs in IC 348: Grain Growth/Settling and Inner Holes?}
\title{24 \micron Detections of Circum(sub)stellar Disks in IC 348: Grain Growth and Inner Holes?}

\author{
James Muzerolle\altaffilmark{1},
Luc\'ia Adame\altaffilmark{2},
Paola D'Alessio\altaffilmark{3},
Nuria Calvet\altaffilmark{4},
Kevin L. Luhman\altaffilmark{5},
August A. Muench\altaffilmark{6},
Charles J. Lada\altaffilmark{6},
George H. Rieke\altaffilmark{1},
Nick Siegler\altaffilmark{1},
David E. Trilling\altaffilmark{1},
Erick T. Young\altaffilmark{1},
Lori Allen\altaffilmark{6},
Lee Hartmann\altaffilmark{4,6},
S. Thomas Megeath\altaffilmark{6}}

\altaffiltext{1}{Steward Observatory, 933 N. Cherry Ave., 
The University of Arizona, Tucson, AZ 85721}
\altaffiltext{2}{Instituto de Astronom\'ia, UNAM, Ap. P. 70-264,
Ciudad Universitaria, 04510, M\'exico, D.F., M\'exico}
\altaffiltext{3}{Centro de Radioastronom\'ia y Astrof\'isica,
Ap. P. 72-3 (Xangari), 58089, Morelia, M\'exico}
\altaffiltext{4}{University of Michigan, 825 Dennison Building,
501 E. University Ave., Ann Arbor, MI 48109-1090}
\altaffiltext{5}{Department of Astronomy and Astrophysics,
The Pennsylvania State University, University Park, PA 16802}
\altaffiltext{6}{Harvard-Smithsonian Center for Astrophysics, 60
Garden St., Cambridge, MA 02138}

\begin{abstract}

We present observations of six late-type members of the young cluster IC 348
detected at 24 \micron with the Multiband Imaging Photometer for {\it Spitzer}
(MIPS).  At least four of the objects are probably substellar.
Combining these data with ground-based optical and near-infrared photometry
and complementary observations with the Infrared Array Camera (IRAC),
we have modeled the spectral energy distributions using detailed
models of irradiated accretion disks.
We are able to fit the observations with models using a range of
maximum grain sizes from ISM-type dust to grains as large as 1 millimeter.
Two objects show a lack of excess emission at wavelengths shortward
of 5.8-8 \micron but significant excess at longer wavelengths,
indicative of large optically thin or evacuated inner holes.
Our models indicate a inner hole of radius $\sim 0.5-0.9$ AU
for the brown dwarf L316; this is the first brown dwarf with evidence
for an AU-scale inner disk hole.  We examine
several possible mechanisms for the inner disk clearing in this case,
including photoevaporation and planet formation.
\end{abstract}


\section{Introduction}

Brown dwarfs have been found in increasingly large numbers, particularly
in young star-forming regions where their larger luminosities and effective
temperatures make them easier to detect.  Observations of young substellar
objects provide important clues to understanding how brown dwarfs
themselves form and, more generally, the nature of the initial mass function.
Of particular interest is the study of circumsubstellar disks.
Since disks are the primary conduits of accretion onto stars and the likely
birthplaces of planetary systems, characterization of disks
around brown dwarfs in comparison to their stellar counterparts
is key to constraining formation mechanisms and exploring the range
of conditions under which planets may form.

Recent investigations by many groups have revealed strong evidence
that disk accretion is an important mechanism in brown dwarf
formation, apparently identical to the processes that operate in
higher-mass T Tauri and Herbig Ae/Be stars.  A remarkable similarity
exists among objects of widely different masses in terms of
both optical and infrared properties associated with the disk accretion
process.  A wealth of permitted emission lines such as HI Balmer
are common in substellar objects; high-resolution spectroscopy
reveals these to be broadened by ballistic infall of gas as expected in
magnetospheric accretion scenarios (e.g. Mohanty et al. 2005;
Muzerolle et al. 2005).
Disk mass accretion rates ($\mdot$) have been measured for brown dwarf disks
using a variety of methods including optical veiling (White \& Basri 2003),
H$\alpha$ profile modeling (Muzerolle et al. 2000, 2003a, 2005),
near-infrared emission lines (Natta et al. 2004), and the Ca II triplet
(Mohanty et al. 2005).  The derived values are extremely
small, typically $\lesssim 10^{-10} \msunyr$, demonstrating that most if not
all disks around low-mass objects must be irradiation-dominated.

Near-infrared excess indicative of
optically thick circumstellar disks has been detected around
many young substellar objects, although often at a marginal level
(e.g. Luhman 1999; Muench et al. 2001; Jayawardhana et al. 2003;
Liu et al. 2003).  Because of the low luminosities of the central sources,
the disk excess produced by irradiation is very small and difficult
to measure at $K$-band and in many cases even at $L$-band.  Thus, observations
at longer wavelengths are necessary to find and characterize
disk emission in detail.  Mid-infrared observations such as ground-based
10 \micron photometry (Apai et al. 2004; Mohanty et al. 2004),
ISO 6.7 and 14 \micron photometry (e.g. Com\'eron et al. 1998, 2000),
and photometry and spectroscopy with the {\it Spitzer} Space Telescope
(Furlan et al. 2005; Luhman et al. 2005a,b) have revealed significant
excess emission around dozens of known young objects near and below
the hydrogen burning limit. 
The shape of the infrared spectral energy distributions (SEDs) of
these brown dwarf disks is qualitatively similar to that of stellar disks
(e.g. Natta \& Testi 2001), with suggestions of a wide range of
substellar disk structure and dust properties (Natta et al. 2002;
Pascucci et al. 2003; Mohanty et al. 2004; Apai et al. 2004).
Further observations such as these are crucial to determining
the properties and complete statistics of brown dwarf disks
in star forming regions and young clusters.

Here, we present detections of very low-mass and substellar members
of the 1-3 Myr-old cluster IC 348 with the 24 \micron channel of
the Multiband Imaging Photometer
for {\it Spitzer} (MIPS).   These data are part of a larger,
more comprehensive {\it Spitzer} GTO imaging study of the cluster
(Lada et al. 2005), a prime target since its stellar and substellar
population has been well-characterized from the ground (e.g. Luhman
et al. 2003).  The MIPS observations provide the longest-wavelength
infrared photometric measurements of brown dwarfs to date, and thus
potentially allow much more stringent constraints on brown dwarf disk
structure than heretofore possible.  Six objects of spectral type M6
and later, near and below the substellar limit, are detected at 24 \micron,
exhibiting emission in excess of expected photospheric levels
and clearly indicative of emission from circum(sub)stellar disks.
In combination with complementary
ground-based optical and near-infrared and {\it Spitzer}
Infrared Array Camera (IRAC) observations,
we present full SEDs from $\lambda \sim 0.8$ to 24 \micron
for these six objects.
We calculate models of irradiated accretion disk emission
including appropriate (sub)stellar parameters,
and examine their capabilities and limitations in describing
the observed SEDs and constraining disk properties such as surface
flaring, inner holes, and dust properties.

\section{Observations}

We have measured 24 \micron photometry from our GTO observations
of IC 348 taken in February 2004 with MIPS (Rieke et al. 2004).
The cluster was mapped
using scan mode with 12 scan legs of length 0.5 degrees and half-array
cross-scan overlap, resulting in a total map size of about 0.5 x 1.5
degrees including overscan.  The total effective exposure time per point
is about 80 seconds.  Data at 70 and 160 \micron were taken simultaneously;
however, we do not consider these here since the sensitivity limits
and strong background emission preclude detection of anything but
the brightest mid-infrared sources.

The 24 \micron images were processed using the MIPS instrument
team Data Analysis Tool, which calibrates the data and applies a distortion
correction to each individual exposure before combining into a final mosaic
(Gordon et al. 2005).  Point source photometry was performed on
the mosaicked image using PSF fitting with {\it daophot}.
The 3-sigma detection limit is on average roughly 1 mJy, as high as 10 mJy in
the central regions of the cluster where the background is strong
and highly structured, and as low as 0.7 mJy in
the lowest-background regions of the map.  In most cases the measurement
errors are dominated by the 10\% uncertainties in the flux calibration.

\section{Analysis}

\subsection{24 \micron properties}

The membership of IC 348 has been well-characterized in various ground-based
optical and near-infrared investigations (Luhman et al. 2003
and references therein).  The low-mass pre-main sequence
population has been confirmed primarily via spectroscopic
indicators of youth such as H$\alpha$ emission, Li absorption,
and various gravity sensitive features.  Here we restrict our analysis
to cluster members near or below the substellar limit, selecting only
objects with spectral types M6 or later.  This cut-off roughly
corresponds to a mass of about $0.1 \; \msun$ at the age of IC 348
(roughly 1-3 Myr) using the pre-main sequence evolutionary tracks
of Baraffe et al. (1998) and the temperature scale of Luhman et al. (2003).

Figure~\ref{bdpos} shows the positions of the 36 known $\ge$M6 members
superposed on the 24 \micron image.  Six of these were clearly detected
and measurable at 24 \micron, while of the remainder, one is an unresolved
companion to a much higher-mass cluster member, and two others are
marginally detected but unmeasurable due to significant
background contamination.  The (sub)stellar properties of the detected
objects are listed in Table~\ref{data}; the IRAC and MIPS fluxes
are listed in Table~\ref{spitz}.  Herein, we refer to these 6 objects as
the low-mass/substellar 24 \micron sample (four objects are below
the substellar limit according to the Baraffe tracks).
The matching to optical/near-infrared
positions in all cases is within about 1 arcsecond, consistent
with the typical pointing accuracy of MIPS.  Confusion with red
background objects is extremely unlikely; in the flux range of
the sample, extragalactic source counts predict roughly 1-10 sources
per square degree (Papovich et al. 2004), resulting in a chance alignment
probability of $\sim 10^{-6}$.  The 24 \micron detections lie primarily
at the outskirts of the cluster, probably reflecting a bias towards
lower-background regions of the map where our detection limits
are lower.

The 24 \micron fluxes of our low-mass sample are all strongly
in excess of expected photospheric levels by factors of 100 or more,
not surprising since the sensitivity limit of our observations is
too high to detect photospheric or weak excess emission from M dwarfs
at the distance of IC 348.
The large mid-infrared excess luminosity implied by our observations
indicates that these objects all harbor significant optically thick disks.
Previous detections of accretion activity on at least 2 of the 6
(L205 and L291; Muzerolle et al. 2003a, Mohanty et al. 2005)
corroborate this conclusion.
Since the flux at 24 \micron is completely dominated by
the excess emission, it offers optimal new constraints on
the properties of disks around very low-mass objects.
%Earlier measurements of brown dwarf disks such as with ISO have probed
%to wavelengths as long as 14 $\mu$m.  These have indicated a range
%of disk surface structure, from evidence of significant flaring
%to evidence of flattening via grain growth and dust settling,
%similar to that seen in T Tauri stars (Natta et al. xxx;
%Mohanty et al. 2004; etc).  24 \micron offers an even more sensitive
%indicator of the disk surface geometry since it probes
%a more distant range of disk radii (roughly 0.5-2 AU??)
%where flaring effects are more significant (D'Alessio et al. xxx, etc).
%It also allows detection of disk with large inner holes by comparison
%with shorter wavelengths, and hence potentially offers a more complete
%disk census in general.  However, we will not address the brown dwarf
%disk fraction in IC 348 here because of the incompleteness of
%the 24 \micron observations, but is investigated at length with the more
%complete data from the Infrared Array Camera (IRAC) in a companion
%paper (Luhman et al. in preparation).

\begin{figure}
\plotone{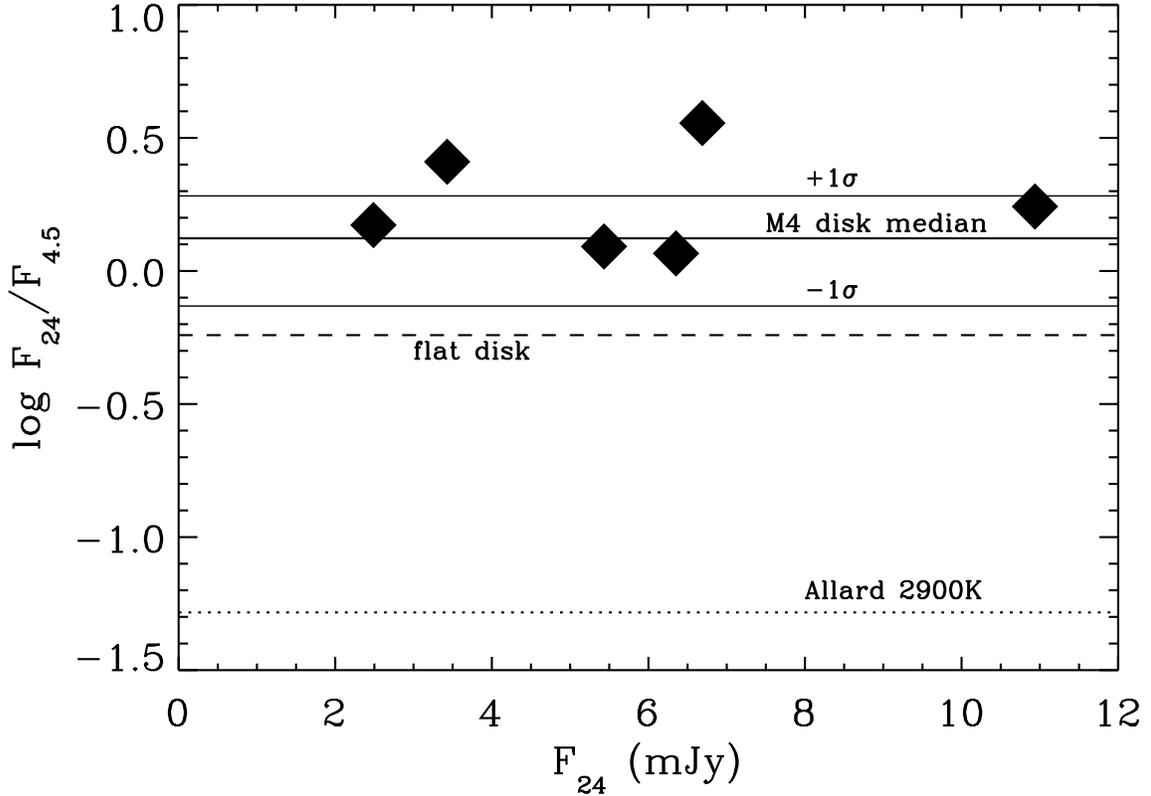}
\caption{The ratio of fluxes at 24 \micron and 4.5 \micron as a function
of the 24 \micron flux for the 6 low-mass/substellar 24 \micron detections
(solid diamonds), all dereddened using $A_V$ determined empirically
from optical photometry and spectral types (Luhman et al. 2003)
and the Mathis (1990) reddening law.
The theoretical flux ratio for an M6.5 photosphere, derived from
an Allard et al. (2001) AMES-COND model for $T_{eff}=2900$ K,
is shown with the dotted line.  The dashed line shows the flux ratio
upper limit for a spectral slope $\lambda F_{\lambda} \propto \lambda^{-4/3}$
corresponding to a geometrically thin, optically thick flat disk,
assuming negligible contribution to the 4.5 \micron flux
from the photosphere.  The solid line shows the median
slope and the dot-dashed lines the $\pm1 \sigma$ values
for $\sim15$ members of IC 348 with spectral types M4-M5 with
excesses indicative of optically thick disks,
as defined in Lada et al. (2005).
\label{fratio}}
\end{figure}

Inferences of disk structure are greatly dependent on theoretical models,
which, as we will show, can be quite degenerate.  However, as an initial guide,
we can examine the observed fluxes in comparison with other higher-mass
objects and simple theoretical expectations.
In Figure~\ref{fratio} we show the ratio of the flux at 24 \micron
to the flux at 4.5 \micron measured from complementary
IRAC observations (Lada et al. 2005; Luhman et al. 2005)
as a function of the 24 \micron flux.
We chose 4.5 \micron as a reference since in most cases
it is the shortest wavelength with relatively unambiguous excess emission
in the lowest-mass objects (though there are a few exceptions; see below).
The low-mass/substellar sample spans a range in the flux ratio of about
0.5 dex, roughly a factor of 50-100 above the theoretical flux ratio of
an M6.5 photosphere.  In comparison, we also plot a flux ratio upper limit
for a flat, optically thick and geometrically thin disk.
All 6 low-mass/substellar 24 \micron detections lie well above this line,
suggesting that the disks are not ``flat" but likely have some amount
of flaring.  However, detailed modeling is required to draw more definitive
conclusions, as we discuss in the next section.

\begin{figure}
\plotone{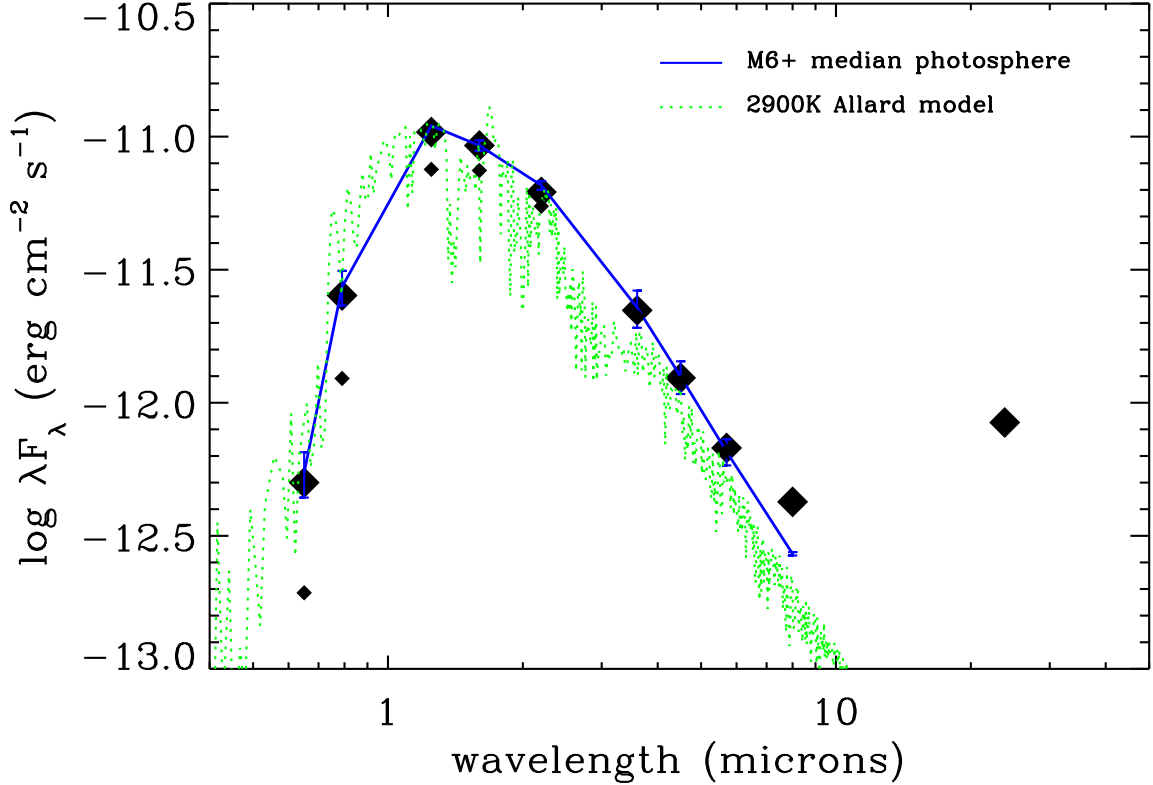}
\caption{Observed SED for the substellar object L316 (diamonds).
Ground-based optical and near-infrared photometry are taken
from Luhman et al. (2003), IRAC data from Lada et al. (2005)
and Luhman et al. (2005).
All fluxes have been dereddened using the observed $A_V$ and
the reddening law of Mathis (1990) (large diamonds); the original reddened
fluxes are also shown (small diamonds).  The median SED and 1-sigma
dispersion for 8 IC 348 members with spectral types later
than M6 which lack infrared excess, normalized to the L316 SED at $J$,
are shown with the solid line and error bars (from Lada et al. 2005).
For further comparison, a photospheric model with
$T_{eff}=2900$ K from Allard et al. (2001) is shown with the dotted
line.  Note the systematically lower flux of this model compared to
the empirical photosphere at the first 3 IRAC bands,
which might cause a false inference of excess emission if one relied
only on the model to represent the intrinsic substellar photosphere.
\label{sedL316}}
\end{figure}

As a further comparison, Figure~\ref{fratio} shows the median
flux ratio of IC 348 late-type stellar members in the range M4-M5
which exhibit excesses indicative of optically thick disks
(Lada et al. 2005).  Four of the six objects in our
low-mass/substellar sample lie within the one-sigma range of
this median ratio, suggesting no significant
differences in disk surface structure for these particular objects.
The other objects, L316 and L30003, exhibit significantly higher flux ratios
than the M4-M5 stellar median.  A closer look at the SED of L316,
shown in Figure~\ref{sedL316}, reveals a lack of excess emission
compared to the photosphere at wavelengths shorter than $\sim$8 \micron.
A similar result is seen in the SED of L30003, but with a cut-off at
a slightly shorter wavelength.
This behavior indicates a significant depletion of small dust grains,
or perhaps no material at all, within the inner $\sim 1$ AU of the disk.
Truncated inner disks associated with the magnetospheric disruption
or dust destruction radius have been previously diagnosed in T Tauri
disks on the basis of the near-infrared excess
spectrum (Muzerolle et al. 2003b), interferometry (e.g. Akeson et al. 2005),
and CO gas emission lines (Najita et al. 2003).
Such truncation radii are typically $<0.1$ AU for typical low-mass
(sub)stellar parameters.  However, the inferred ``hole" sizes
for L316 and L30003 are much larger, more akin to ``transition" disks
found around higher mass stars such as CoKu Tau/4 (D'Alessio et al. 2005)
and TW Hya (Calvet et al. 2002),
and thus require a different creation mechanism.
%One possibility is the effect of planetesimal growth and/or planet formation,
%which if true would be the first such evidence in substellar objects.
We examine several possibilities in detail in section 4.2.

\subsection{SED modeling}

We next compare the observed infrared emission of each of
the low-mass/substellar objects in our sample 
with detailed vertical structure models of irradiated accretion disks.
The assumptions, equations and description of the method used
to calculate the disk structure and emergent intensity are given in
D'Alessio et al. (1998, 1999, 2001); we defer the details to these papers.
In brief, the disks are assumed to be in steady-state,
with dust and gas well-mixed and thermally coupled and the mass accretion
rate uniform throughout the entire disk.  We adopt a viscosity parameter
$\alpha=0.01$, as is typically assumed for T Tauri disks (but not
well-constrained; see e.g. Hartmann et al. 1998).  The dust grain size
distribution is given by the standard power law $n(a)\sim a^{-3.5}$
(Mathis, Rumpl, \& Nordsieck, 1977), with minimum grain size
$a_{min}=0.005$ \micron and maximum grain size varied to be either
$a_{max}=0.25$ \micron (consistent with ISM dust; D'Alessio et al. 1999)
or $a_{max}=1$ mm (the maximum grain size of the model that best fits
the median SED of classical T Tauri stars in Taurus; D'Alessio et al. 2001)
as a preliminary exploration of dust grain growth within these disks.  
The inner disk radius $R_{wall}$ for a ``\emph{normal}" disk
is given by the dust sublimation radius $R_{sub}$ where $T_{sub}\sim1400$ K.
In a few cases, we also consider larger inner radii (``\emph{truncated}")
as the observed near-infrared emission dictates.
We include the emission of a vertical wall at the inner edge,
where the dust has been heated by frontal irradiation from
the central object (see Muzerolle et al. 2003b and D'Alessio et al. 2005
for details of the treatment).  
If an object is actively accreting and has a dust sublimation radius
larger than the corotation radius (as is usually the case),
a pure gaseous disk must be present interior to $R_{sub}$.
Since emission from this component is negligible at low accretion rates
(Muzerolle et al. 2004), we neglect it in our models.
%All models presented here implicitly assume that the dust sublimation radius
%occurs {\it outside} the magnetospheric truncation radius,
%which need not be the case; note that if the dust disk is truncated at
%the magnetospheric radius the temperature of the inner wall would be
%lower than 1400 K.
The maximum disk radius is fixed at $100$ AU; no constraints on disk radii
are available for the objects under study here, but in any case the model
SEDs are not very sensitive to this parameter in the wavelength range
we consider.

\begin{figure}
\plotone{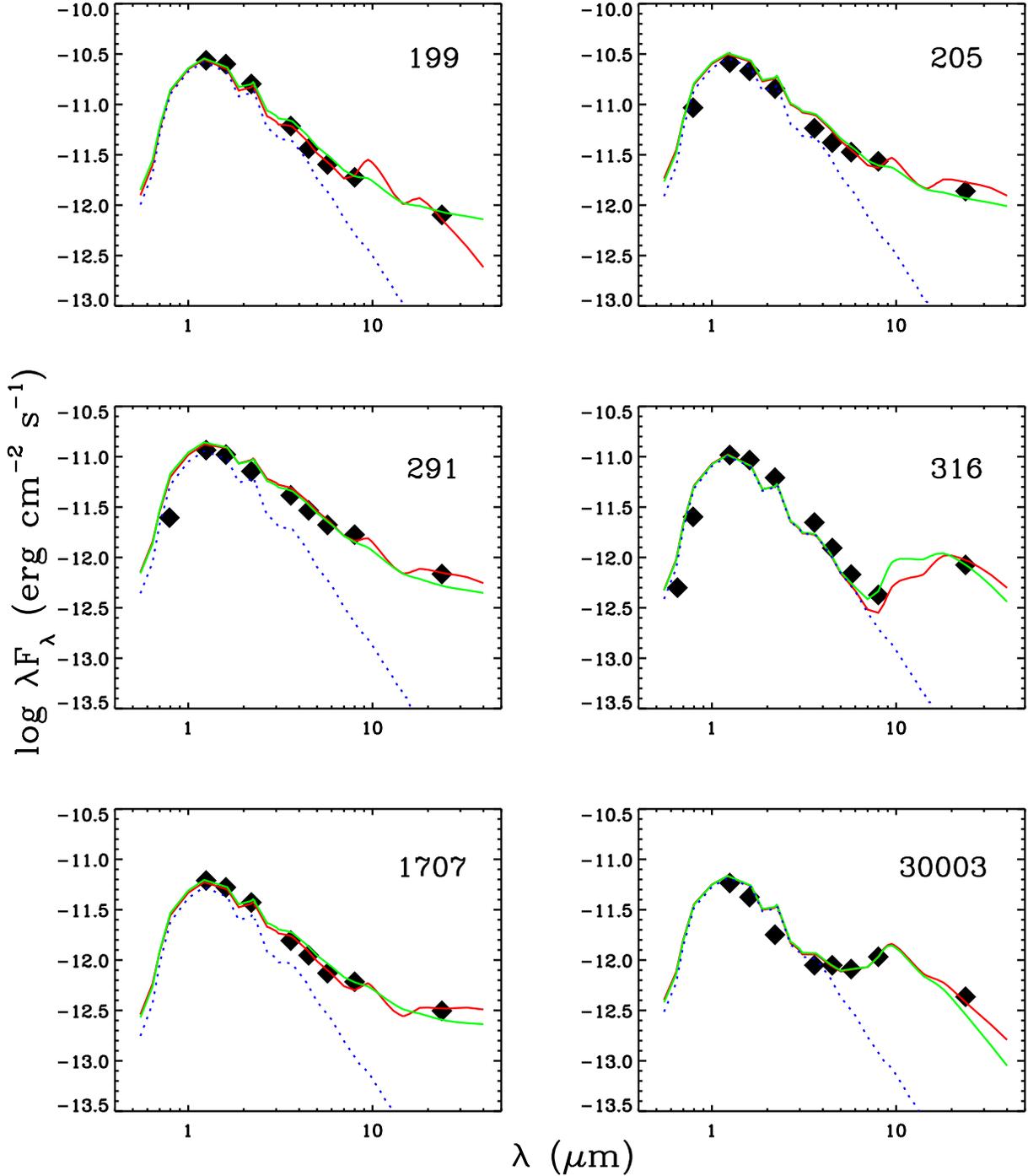}
\caption{Observed SEDs for all 24 \micron detections with spectral types
M6 or later (squares).  Ground-based optical and near-infrared photometry
from Luhman et al. (2003) and Luhman et al. (2005, in preparation),
IRAC data from Lada et al. (2005) and Luhman et al. (2005).
All fluxes have been dereddened using the observed $A_V$ and
the reddening law of Mathis (1990).  Best-fit disk models with
maximum grain size $a_{max}=0.25$ \micron and 1 mm are shown
with the red and green solid lines, respectively.  These models include
a photospheric component using the theoretical models of Allard et al. (2001)
appropriate for the spectral type of each object, smoothed versions
of which are shown separately with the blue dotted lines.
\label{sedism}}
\end{figure}

For comparison to the observed SEDs, we adopted (sub)stellar
parameters as shown in Table~\ref{data} and a distance to IC 348 of
315 pc.  The observed fluxes were corrected
for extinction according to the Mathis (1990) reddening law
given the empirical A$_V$ values shown in Table~\ref{data}.
For each object, we explored a range of inclination angles
to the line of sight, and three mass accretion rates
($10^{-12}$, $10^{-11}$, and $10^{-10}$ $\msunyr$) which are in
the typical range for young very low mass stars and brown dwarfs
(Muzerolle et al. 2003a, 2005).  The height of the inner disk wall
was then adjusted to fit the observed flux in the 2-8 \micron range.
Table~\ref{models} shows the input parameters of the best fit models for each
observed SED and the resultant physical parameters of the wall and disk.
The SEDs are shown in Figure~\ref{sedism}.

\begin{figure}
\plotone{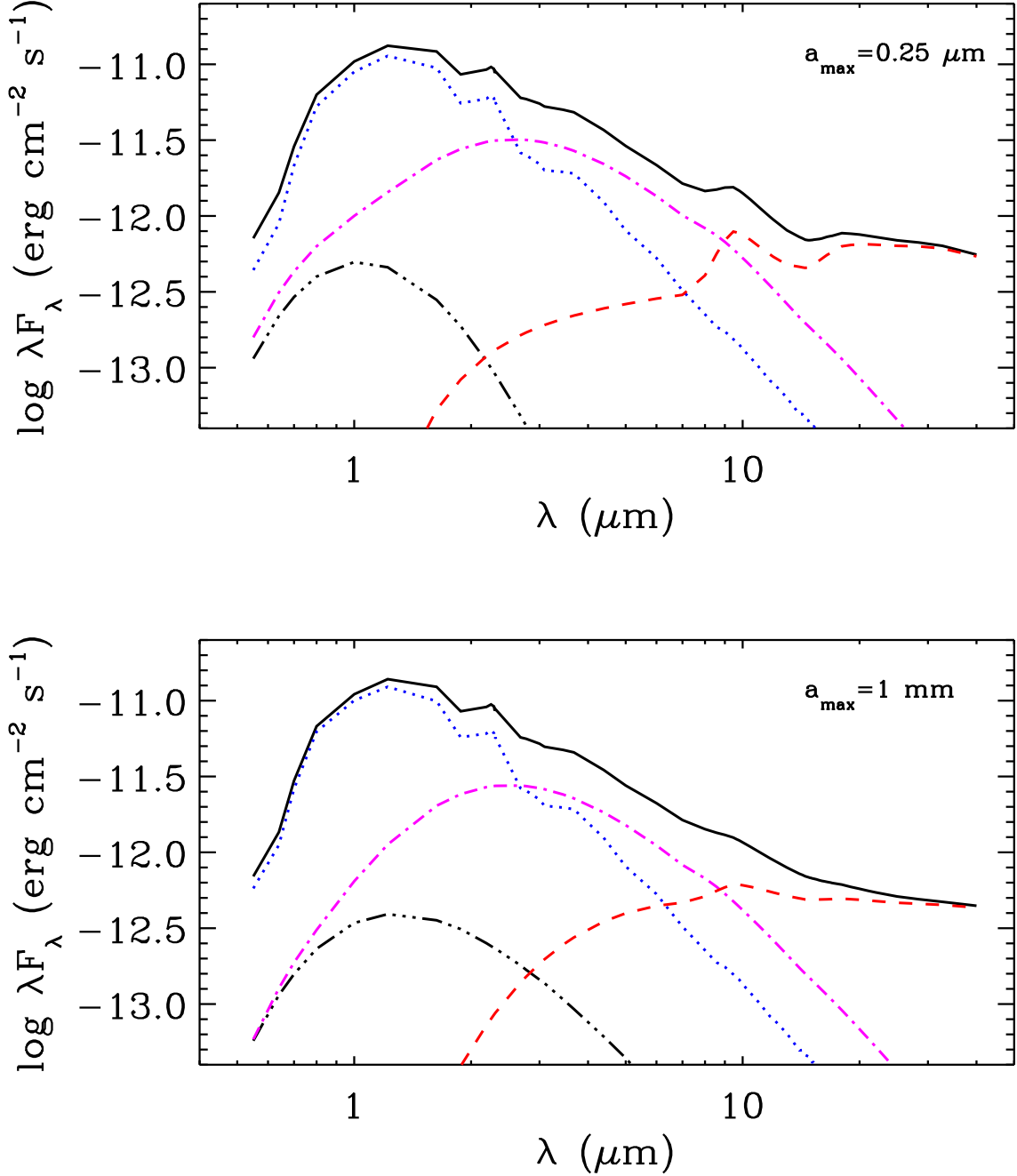}
\caption{Brown dwarf disk models for L291 parameters:
(top) for $a_{max}=0.25$ \micron; (bottom) for $a_{max}=1$ mm.
Blue dotted lines are Allard et al. (2001) model atmospheres
with $T_{eff}=2900$ K, $log(g)=3.5$;
red dashed lines are the emission from the disk surface;
magenta dash-dotted lines are the emission from the inner disk wall;
black dash-dot-dot lines are the contribution from scattered light;
solid black lines are the total emission.
\label{model}}
\end{figure}

For objects L199, L205, L291, and L1707, ``\emph{normal}" disk models
with a range of $a_{max}$ are capable of reproducing the observed SEDs.
The emission arising from the inner disk wall
is required in order to match the fluxes in the 2-8 \micron range,
while the outer disk is only responsible for the emission
redward of  $\sim 10$ \micron; Figure~\ref{model} shows
the relative contributions of the various components of a typical model.
We find that a disk with $\mdot=10^{-11} \msunyr$,
$a_{max}=0.25$ \micron, and an inclination angle of $60^\circ$ provides
reasonable fits for L205, L291, and L1707.  In the case of L199,
we can only match the observed SED with $a_{max}=0.25$ \micron
for a relatively extreme combination of pole-on disk with small
accretion rate $\mdot=10^{-12} \msunyr$.
Disks with larger maximum grain sizes up to $a_{max}=1$ mm
can also fit the observed SEDs if the accretion rates are increased by
one or more orders of magnitude and the inclination angles
are changed significantly (nearly pole-on orientations for L205, L291,
and L1707 and $60^\circ$ for L199).  To fit the near-infrared fluxes
we also have to change the wall height to compensate for the change in
model flux as a result of orientation effects (i.e., the wall emission
decreases as the inclination becomes more pole-on because of our assumed
vertical geometry).  The adopted heights are shown in
Table~\ref{models}, along with the expected height given
$a_{max}=0.25$ \micron in the wall and equations 5 and 6
from Muzerolle et al. (2004).  Figure~\ref{model} compares
models with the two maximum grain size extrema; the overall disk flux
is similar in both models owing to the adjustments in $\mdot$ and inclination,
but the 10 and 20 \micron silicate features, produced in the atmosphere
of the outer disk, are markedly weaker for $a_{max}=1$ mm (see also
D'Alessio et al. 2001).
Spectra of these features are obviously important to discriminate
between the two cases.

\begin{figure}
\plotone{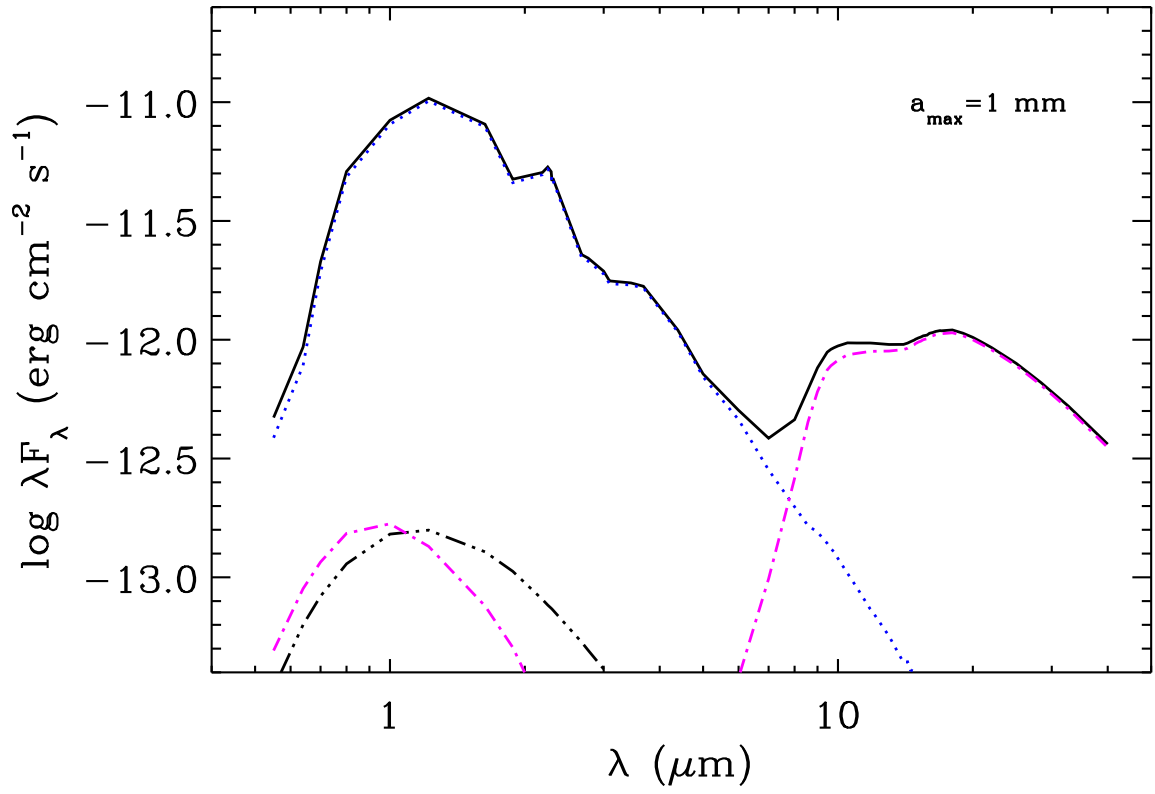}
\caption{Same as in Figure~\ref{model} with $a_{max}=1$ mm for the case
of a large ($\sim1$ AU) inner disk hole.
\label{model_hole}}
\end{figure}

As already mentioned, objects L316 and L30003 exhibit a lack of excess emission 
at wavelengths smaller than $5.8-8$ \micron, indicating a lack of hot dust.
We could not obtain a good fit with any ``normal" disk model,
thus, we have calculated the emission from a disk \emph{truncated}
at radii larger than the dust sublimation radius,
following the treatment of Muzerolle et al. (2003b)
and D'Alessio et al. (2005).  The inner wall in this case is
the transition between an optically thin or completely evacuated hole
and an optically thick outer flared disk which contributes
to the SED only at longer wavelengths, generally $>10$ \micron.
Again we assume that $a_{max}=0.25$ \micron;
the maximum grain size is varied between 0.25 \micron and 1 mm
for the outer disk only.  Figure~\ref{model_hole} shows
the emission components for one such model run for L316.
The emission redward of 8 \micron
is dominated entirely by the vertical wall, which is located near
1 AU and hence has a significantly lower surface temperature
($\sim200$ K) than the dust sublimation point.
We have set $\dot{M}=10^{-12} \; M_\odot$ yr$^{-1}$ for models of
both L316 and L30003, but observational constraints are lacking.
This level is probably an upper limit for L316 since its small
H$\alpha$ equivalent width (3 \AA) indicates either very weak or
completely blocked accretion onto the central object.
%In any case, the viscous heating from such a small accretion rate
%is negligible and has no effect on the SED.

\section{Discussion}

\subsection{Grain Growth and Dust Settling?}

Comparisons of disks around low-mass stars and brown dwarfs
can in principle provide clues to differences or similarities
in formation mechanisms, as well as put limits on the possibility
of planet formation.  Mid-infrared measurements probe dust emission
from the terrestrial zone of disks and provide a useful point of comparison
for investigating bulk dust properties.  Of particular interest
has been the detection of infrared signatures of disk evolution
via grain growth and settling.
Evidence for such processes has been inferred for ``mature" T Tauri disks
at ages of up to 5-10 Myr (Calvet et al. 2005a; Sicilia-Aguilar et al. 2005).
As mentioned in the introduction, previous measurements of thermal dust
emission from a small number of brown dwarfs
have indicated a range of disk properties, from surface flaring
and dust grain sizes typical of most 1 Myr-old T Tauri stars
to flatter disks with evidence for significant
dust grain growth and settling towards the disk midplane.
These inferences are based primarily on observed fluxes out to 14 \micron
and rough silicate emission strengths from narrow-band 10 \micron imaging
(Apai et al. 2004; Mohanty et al. 2004; Sterzik et al. 2004).
Our 24 \micron observations potentially offer an even more
sensitive indicator of the disk surface geometry since they probe
a more distant range of disk radii (roughly 0.5-1 AU for a central
substellar object) where the disk height is up to 30\% greater
(D'Alessio et al. 1999).

The flux ratios shown in Figure~\ref{fratio} suggest that the disks
in our sample are not flat but must have some amount of flaring,
to a degree not dissimilar
to that seen in the low-mass stellar population of IC 348 in general.
Lada et al. (2005) compare stellar SEDs as a function of spectral type
and find that the latest-type stars ($>$M4) with disks exhibit steeper
spectral slopes than seen in earlier-type stars with disks,
possibly indicating a higher frequency of dust settling around lower-mass
objects.  However, our disk model comparisons indicate
that we cannot easily constrain the dust properties
for most of the low-mass/substellar sample.
Without additional information such as the mass accretion rate or
the 10 and 18 \micron silicate features, the observed excess emission out
to 24 \micron can be explained by a wide range of maximum dust grain sizes,
from ISM-type to as large as 1 mm.
Observational constraints on the disk mass accretion rates are
available only for L205 ($\mdot=10^{-10} \msunyr$; Muzerolle et al. 2003a)
and L291 (an upper limit of $\mdot<10^{-11} \msunyr$; Mohanty et al. 2005).
These would seem to argue against the $a_{max}=0.25$ \micron model for L205
since the model accretion rate is a factor of 10 lower.
If this is the case, we may be seeing evidence of grain growth
in the L205 disk.  For object L199,
the $a_{max}=0.25$ \micron model only fits
for rather extreme values of $\mdot$ (small) and inclination (pole-on),
which may not be realistic particularly since pole-on orientations
should be statistically rare.  Similarly, the $a_{max}=1$ mm models
may not hold for L291 and L1707 since they also require pole-on orientations.

One suggestion of dust settling may be seen in the vertical heights
of the inner disk wall component of the models.  The heights required
to match the short-wavelength excess of objects L199 and L205 are factors
of 1.5-2 smaller than the expected heights derived from equation 6 of
Muzerolle et al. (2004).  It is possible that grain growth and settling
in the inner disk can result in smaller than expected wall heights.
However, the exact values constrained by our models are limited by
uncertainties in the level of the underlying photosphere
(see Table~\ref{models}) as well as the uncertain structure of the inner wall.
Models incorporating the wall with the outer disk in a self-consistent
manner, including the effects of dust settling to the disk midplane,
must be explored before definitive conclusions can be made.
Again, mid-infrared spectra covering the silicate features are
also crucial observational constraints that should be obtained.
We further caution that all the ``normal" disk models presented here
implicitly assume that the wall radius is at the dust sublimation point.
Given the cool and faint radiation fields of very low-mass objects,
the sublimation radius is quite small (Table~\ref{models}) and may
in fact be smaller than the radius at which the (sub)stellar
magnetosphere truncates the gas disk.  If this is the case,
the dust edge should then be at the magnetospheric radius and can have
a dust temperature less than 1400 K, yielding weaker near-infrared excess
emission (a lower temperature might also provide a better fit
to the spectral slope at the IRAC bands in L205).
Since estimates for the magnetospheric truncation radii
are highly uncertain, we leave this complication for a future study.

We must also point out that all of the models we consider here assume
the same constant value of the viscosity parameter $\alpha$.
The model SED at $\lambda > 10$ \micron can be modified
significantly if $\alpha$ is changed by just an order of magnitude.
There is no theoretical nor observational impediment, at this time,
to adopting different values of $\alpha$.  For smaller values,
the disk density will increase assuming a constant $\mdot$,
since $\Sigma\propto\frac{\dot{M}}{T_c\alpha}$,
where $T_c$ is the temperature at the disk midplane.
If the density increases, the height of the irradiation surface,
defined as the height where the mean optical depth to the (sub)stellar
radiation is unity, also increases.  The higher the irradiation surface,
the larger the fraction of incident (sub)stellar flux is intercepted and
reprocessed by the disk.  On the other hand, the mean opacity to
the (sub)stellar radiation is smaller with increasing $a_{max}$
from ISM-like values to millimeter sizes.  Thus, different combinations
of $\mdot$, $\alpha$, and $a_{max}$ can fit the same SED.
A full exploration of the effects on the structure of disks
around very low mass objects produced by
varying $\alpha$ are beyond the scope of this work.
In any case, 10 \micron spectra and millimeter observations
are required for a more rigorous test of the grain growth hypothesis
for objects L199 and L205.  Models including the effects of dust settling
to the disk midplane should also be explored.

%These results represent the few IC 348 low-mass/substellar
%disks we detect at 24 \micron, and thus may only be the tip of the iceberg.
%We cannot rule out the presence of other disks with significant
%dust settling in IC 348.  Such disks would in general display weaker
%24 \micron excesses below our detection threshold.  The brown dwarf disk
%fraction in IC 348, based on IRAC measurements alone (in most cases
%sensitive to brown dwarf photospheric levels at this distance), is about
%40\% (Luhman et al. 2005).  Of these, 33\% (5 objects) are detected
%at 24 \micron.  Assuming the typical 24/4.5 \micron ratio shown
%in Figure~\ref{fratio} and a median background sensitivity limit of
%1 mJy, we estimate that four of the nondetections
%with IRAC excesses should have been detected; unfortunately, three of these
%are located on extremely high backgrounds and one is severely confused
%with the Airy pattern of a bright source.  Thus, we cannot make any
%meaningful statements about sources not detected in this region.

%13 IRAC excesses M6 and later out of 35:
%M6 excesses: L205, 325, 622, 725 (out of 11)
%M6+ excesses: 478, 415, (L316), 407, L291, 703, 468, 603, 690, 738, 4044
 %(out of 24) plus 199, L291 from this paper
%
%typical [4.5]-[24] ~ 3.9
%not on bright background at 24 (predicted 24 based on 4.5 mag in paren.):
% 468 (8.9, marginally detected?), 407(9.8, marg?), 690(10.4), 478(9.6),
% 725(12.1), 738(11.0), 4044(?)
%on/near bright background:
% 622(9.8), 603(10.0), 415 (8.5, marg?), 325 (8.2, v. bad back.) 

\subsection{Inner Disk Clearing}

Two objects, L316 and L30003, show evidence for inner disk clearing
beyond what could be explained by magnetospheric truncation or
the dust destruction radius alone.  L316 is the first young brown dwarf
known to exhibit a significant inner disk ``hole".
This result has potentially significant impact on our understanding
of disk clearing mechanisms since the low object masses provide
stringent constraints on theoretical models previously proposed
to explain similar ``transition" disks around higher-mass young stars.
We discuss two possibilities, photoevaporation by the UV flux of 
the central object and clearing via grain growth or planet formation.

\subsubsection{Photoevaporation}

Clarke et al. (2001), building on the pioneering work of Hollenbach et al.
(1994), first proposed a photoevaporative wind as a means of
producing inside-out clearing of circumstellar accretion disks.
In brief, the model treats mass loss from the disk surface as a result
of photoevaporation caused by the impinging UV radiation of the central
star.  This loss occurs beyond a characteristic ``gravitational radius"
($R_g$) that depends on the mass of the central object and the sound speed
of the photoionized gas.  As the disk viscously evolves with time
its mass accretion rate decreases steadily; once the accretion rate drops
below the photoevaporative mass loss rate, the disk experiences a net loss
of material outside $R_g$.  A gap quickly forms at $R_g$ ($\sim 5-10$ AU
for typical T Tauri stellar parameters), and the continuing mass loss
prevents replenishment by accreting material from outside.
The inner disk, now decoupled at $R_g$, drains onto the star in
a viscous time (a few times $10^5$ yrs for typical T Tauri stars),
leaving a hole in the disk of size $R_g$.
The outer disk then continues to evaporate away more slowly.

The main difficulty with this model has been the origin of sufficient UV flux.
Matsuyama et al. (2003) showed the excess hot continuum produced by
typical T Tauri accretion shocks does not produce enough UV flux,
particularly when the decrease in mass accretion
rate with time expected by viscous evolution is folded in.
We can now place more stringent constraints on the timescales for
this process to operate by inference from the very low-mass objects
which show inner disk clearing.
In particular, the hole radius we infer for object L316 with
the $a_{max}=1$ mm model is similar to the gravitational radius
$R_g = GM_*/a^2 = 0.5$ AU, where $a$ is the sound speed of the ionized
gas with assumed $T=10^4$ K.  Ruden (2004) gives useful scaling
relations for the photoevaporative mass loss
using a framework similar to that of Matsuyama et al.,
assuming that the accretion shock is the only source of significant
EUV photons and emits blackbody radiation with
characteristic temperature 15,000 K.  Most applicable for our purposes is 
the characteristic timescale for disk removal,
$t_e = 3.4 \times 10^7 \, yr \; (M_{d0}/10^{-2}\msun)^{2/3} \,
(M/\msun)^{-5/6} \, (\alpha/10^{-2})^{-1/3}$, and the total mass removed,
$\Delta M_w = 0.3 \, M_J \; (M_{d0}/10^{-2}\msun)^{2/3} \, (M/\msun)^{7/6}
\, (\alpha/10^{-2})^{-1/3}$.
% (the constants have been modified slightly
%from Ruden to account for the smaller radii of our objects). 
We adopt an initial disk mass $M_{d0}=10 \; M_J$ as a plausible
upper limit{\footnote{This gives a ratio $M_{disk}/M_* \sim 0.1$,
at the high end of the range measured in T Tauri stars and close
to the limit for gravitational instability.
The only published disk mass
measurements for substellar objects are from Klein et al. (2003),
who estimated values in the range 0.4-2.4 and 1.7-5.7 $M_J$ for
the brown dwarfs CFHT-BD-Tau 4 and IC 348 613, respectively.}},
and a viscosity parameter
$\alpha=0.01$.  With these assumptions, we estimate
that a total of only $\sim 4.7 \; M_{\oplus}$ can be removed
via a photoevaporative wind in about 300 Myr, some 300 times longer
than the age of the object.  Smaller values of $\alpha$ will only increase
the evaporative timescale; smaller initial disk masses will decrease
the timescale but also vastly decrease the mass removed.
Thus, we conclude that a photoevaporative wind generated by UV radiation from
a substellar accretion shock cannot have produced the inner disk
hole we see in L316.  In fact, there may be no accretion shock at all
on L316 if accretion has indeed shut off, as its H$\alpha$ equivalent
width seems to suggest (though a resolved line profile would
provide a better indicator).  An alternative source of the UV flux may arise
from the chromosphere; however, it would need to provide a steady
flux of $\Phi \gtrsim 10^{40}$ s$^{-1}$ (Ruden 2004),
which is difficult to imagine for such a low-mass object.
However, measurements of the chromospheric UV flux in young objects
of any mass are lacking.

\subsubsection{Planetesimal growth and/or planet formation}

Significant grain growth and settling in the inner disk
may lead naturally to rapid formation of meter or kilometer-sized
planetesimals.  Models of dust coagulation in protoplanetary disks predict
growth at a rate proportional to the square of the orbital period,
with faster timescales in the inner disk because of the higher velocities
(e.g. Weidenschilling 1997).  Recent simulations along these lines
by Dullemond \& Dominik (2005) demonstrated that small grains can be removed
quickly from the inner regions of T Tauri disks, so long as they are not
replenished.  In fact, the process appears to act too quickly,
in less than 1 Myr, and replenishment by aggregate fragmentation
from high speed collisions was suggested by Dullemond \& Dominik
as a way to extend disk lifetimes to agree with observations.
Larger scale effects associated
with the dissipation of the gas in the disk may also cause planetesimal
orbits to shift and create additional collisions since the
system will never reach a static state until the gas is gone.
Other proposed mechanisms for the removal of small dust grains in disks
around low-mass stars, such as corpuscular wind drag (Plavchan et al. 2005),
only apply once the gas has completely dispersed.
Therefore, the attribution of cleared inner disks to clearing by
planet formation brings its own set of complications.
It may mean a system of planetesimals akin to
the asteroid belt within the cleared inner zone at a time away
from major collision events, or it may
suggest the very rapid formation of a massive planet within a few AU
of the central object, as has been suggested for T Tauri stars with
similar evidence for inner disk holes such as TW Hya (Calvet et al. 2002),
GM Aur (Rice et al. 2003; Calvet et al. 2005b),
and CoKu Tau/4 (D'Alessio et al. 2005).

The case of L316 may illustrate the latter of these possibilities.
Accretion has probably ceased altogether,
and if so there is likely little or no gas
inside the hole.  In order to prevent accretion from the outer disk
from proceeding through the hole onto the star, dynamical perturbation
from a planetary-mass object might be required.
Hydrodynamic simulations by Quillen et al. (2004)
have shown that planetary companions around young stars can disrupt
the inner disk and prevent further accretion from the outer disk
if they are sufficiently massive.  They estimate this lower limit
in the case of the inner disk hole-bearing T Tauri star CoKu Tau/4
using the gap-formation criterion that the ratio of the planetary
and primary masses $q \ge 40$ Re$^{-1}$, where the Reynolds number $Re$
is a function of the disk viscosity (Nelson et al. 2000).
For the $\alpha$-disk prescription, $Re = {\alpha}^{-1}
(v_c/c_s)^2$, where $v_c$ is the Keplerian velocity of a particle
at a given radius in the disk and $c_s$ is the sound speed at the same radius.
In the case of object L316, we adopt an inner hole size of 1 AU,
the temperature at the disk edge $T \sim 200$ K, and the Keplerian
velocity at the edge given the mass of L316.  Assuming $\alpha = 0.01$,
we then derive a rough lower limit $M_p \gtrsim 0.8 \; M_J$
for the planetary mass necessary to maintain the observed hole.

The initial disk around L316 most likely did not have enough material
within 0.5-1 AU to form a planet this large.  We have simulated
a possible initial disk using a standard model with $\alpha = 0.01$,
inner edge at the dust sublimation radius, and a much larger accretion rate
of $10^{-9} \; \msunyr$ (a plausible initial value based on
the observed range of accretion rates in brown dwarfs;
Muzerolle et al. 2005).  The resulting model contains about
2-3 $M_{\oplus}$ of total dust+gas inside 0.5 AU.
However, $\alpha$ is not known; values of $10^{-4} - 10^{-3}$
are theoretically plausible and would give smaller lower limits
to the planetary mass necessary to maintain the inner hole,
$M_p \gtrsim 2.5-25 \; M_\oplus$, as well as more mass in the initial disk.
A lower $\alpha$ could for instance be indicative of a disk in which
the midplane is inactive (``dead zone"; Gammie 1996), with very low
or zero equivalent $\alpha$, and accretion occurs only in an active layer
with larger $\alpha$.  The same initial disk model with these smaller
values of $\alpha$ results in disk masses within 0.5 AU of up to 60 $M_\oplus$.
Thus, a sufficient reservoir of material could have been present around
L316 initially for this scenario to hold.  However, the mean age of IC 348
members then requires a fast planet formation timescale of
a few Myr or less.  The surface densities of the initial disk models
are too low by several orders of magnitude for the disk instability
model of Boss (1997) to apply.  More work needs to be done
to evaluate whether the traditional models of core formation
(or even ``super-Earth'' formation) via planetesimal collisions
can build an object in the mass range we estimate in the time required.

\section{Summary}

We have reported on the detection at 24 \micron of dust emission
from 6 very low mass or substellar members of the 1-3 Myr-old cluster
IC 348.  The amount of excess emission in all cases is consistent
with an origin in optically thick disks.  In combination with ground-based
and {\it Spitzer} IRAC photometry, we find that the SED shapes of
four objects are similar to those of most of the disks around M stars
in the same cluster, indicating similar flared disk geometries.
Models of irradiated accretion disks with maximum grain sizes no
larger than 1 millimeter fit the SEDs, however, in most cases we cannot
confidently rule out smaller maximum grain sizes.
Measurements of mass accretion rates,
spectra of the 10 \micron silicate feature, and/or millimeter-wave
observations are needed to better constrain the models.  The two objects
with dissimilar SEDs exhibit a lack of excess emission shortward
of 5-8 \micron indicative of evacuated inner holes.  Our models
constrain the size of the inner hole in object L316 to $R_{in} \sim$ 0.5-1 AU.
This is the first brown dwarf seen to exhibit evidence for a transition disk
with an AU-scale inner hole.
Processes related to planet formation may have contributed to
the clearing of the inner disk, either via planetesimal coagulation
to asteroid-size bodies or formation of a single ``super-Earth''.
If true, this suggests that planet formation can occur around objects
across the full range of stellar and substellar masses, and puts
strong constraints on the formation timescale of a few Myr or less.
Further models of grain growth and planetesimal coagulation
in such environments need to be advanced in order to assess
these possibilities, or whether a different clearing mechanism is required.

\acknowledgements

This work is based in part on observations made with the {\it{Spitzer Space
Telescope}}, which is operated by the Jet Propulsion Laboratory,
California Institute of Technology under NASA contract 1407.
Support for this work was provided by NASA through Contract Number
960785 issued by JPL/Caltech.  L. A. acknowledges support from CONACyT
grant 172854, and P. D. acknowledges grants from DGAPA, UNAM, and CONACyT,
M\'exico.  K. L. was supported by grant NAG5-11627 from the NASA
Long-Term Space Astrophysics program.

\begin{deluxetable}{lccccccccccc}
%\setlength{\tabcolsep}{0.04in}
%\tabletypesize{\scriptsize}
\tablewidth{0pt}
\tablecaption{The Low-Mass/Substellar 24 \micron Sample \label{data}}
\tablehead{
\colhead{ID} &
\colhead{$\alpha$(J2000)} &
\colhead{$\delta$(J2000)} &
\colhead{Sp. Type} &
\colhead{$A_{V}$} & \colhead{$M_*$ ($\msun$)} & \colhead{$R_*$ ($\rsun$)} &
\colhead{$L_{\rm bol}$ ($\lsun$)}}
\startdata
L199$^a$ &  03:43:57.22 & 32:01:33.9 & M6.5 & 9.2 & 0.08 & 1.15 & 0.087\\
L205$^b$ & 03:44:29.80 & 32:00:54.6 & M6 & 2.3 & 0.1 & 1.15 & 0.095\\
L291$^c$ & 03:44:34.05 & 32:06:56.9 & M7.25 & 1.0 & 0.05 & 0.74 & 0.032\\
L316$^c$ & 03:44:57.73 & 32:07:41.9 & M6.5 & 1.2 & 0.075 & 0.71 & 0.033\\
L1707$^a$ & 03:43:47.63 & 32:09:02.7 & M7 & 1.0 & 0.06 & 0.53 & 0.017\\
L30003$^a$ & 03:43:59.17 & 32:02:51.3 & M6 & 8.0 & 0.1 & 0.54 & 0.021
\enddata
\tablecomments{Object IDs and properties have the following
references: $^a$Lada et al. in preparation; $^b$Luhman (1999);
$^c$Luhman et al. (2003).}
\end{deluxetable}

\begin{deluxetable}{lccccc}
%\setlength{\tabcolsep}{0.04in}
%\tabletypesize{\scriptsize}
\tablewidth{0pt}
\tablecaption{{\it Spitzer} Fluxes \label{spitz}}
\tablehead{
\colhead{ID} & \colhead{$F_{3.6}$} & \colhead{$F_{4.5}$} &
\colhead{$F_{5.8}$} & \colhead{$F_8$} & \colhead{$F_{24}$}}
\startdata
L199 & 4.87 $\pm$ 0.28 & 4.08 $\pm$ 0.18 & 3.92 $\pm$ 0.22 & 3.84 $\pm$ 0.12 & 5.58 $\pm$ 0.24\\
L205 & 6.12 $\pm$ 0.06 & 5.81 $\pm$ 0.13 & 6.06 $\pm$ 0.31 & 6.79 $\pm$ 0.18 & 10.58 $\pm$ 0.32\\
L291 & 4.73 $\pm$ 0.09 & 4.25 $\pm$ 0.10 & 3.90 $\pm$ 0.14 & 4.35 $\pm$ 0.22 & 5.35 $\pm$ 0.23\\
L316 & 2.53 $\pm$ 0.02 & 1.79 $\pm$ 0.06 & 1.25 $\pm$ 0.08 & 1.02 $\pm$ 0.16 & 6.57 $\pm$ 0.22\\
L1707 & 1.79 $\pm$ 0.03 & 1.62 $\pm$ 0.02 & 1.38 $\pm$ 0.07 & 1.58 $\pm$ 0.06 & 2.45 $\pm$ 0.09\\
L30003 & 0.75 $\pm$ 0.04 & 1.03 $\pm$ 0.03 & 1.30 $\pm$ 0.12 & 2.27 $\pm$ 0.13 & 3.06 $\pm$ 0.35
\enddata
\tablecomments{Fluxes in mJy.  MIPS 24 \micron flux uncertainties are
the formal errors from the PSF-fitting procedure, and do not include the 10\%
calibration uncertainty.}
\end{deluxetable}

\begin{deluxetable}{lcccccc}
%\tabletypesize{\small}
\tablecaption{Disk Model Parameters \label{models}}
\tablewidth{0pt}
\tablehead{
\colhead{Object} & \colhead{$T_{wall}$ (K)} & \colhead{log $\mdot$} &
\colhead{$\cos(i)$} & \colhead{$R_{wall}$ ($R_*$)} &
\colhead{$H_{wall}$ ($R_*$)} & \colhead{$H_{wall}^{pred}$ ($R_*$)}}
\startdata
&&& $a_{max} = 0.25$ \micron \\
L199 & 1400 & -12 & 0.9 & 4.5 & 0.1 & 0.22\\
L205 & 1400 & -11 & 0.5 & 4.7 & 0.2 & 0.23\\
L291 & 1400 & -11 & 0.5 & 4.1 & 0.4 & 0.25\\
L316 & 200 & -12 & 0.8 & 276.2 & 8.0 & \nodata \\
L1707 & 1400 & -11 & 0.5 & 4.1 & 0.2 & 0.26\\
L30003 & 500 & -12 & 0.5 & 45.5 & 2.0 & 1.3\\
\hline
&&& $a_{max} = 1$ mm \\
L199 & 1400 & -10 & 0.5 & 4.5 & 0.1 & 0.22\\
L205 & 1400 & -10 & 0.8 & 4.7 & 0.2 & 0.23\\
L291 & 1400 & -10 & 0.9 & 4.1 & 0.6 & 0.26\\
L316 & 250 & -12 & 0.5 & 165.4 & 8.0 & \nodata \\
L1707 & 1400 & -10 & 0.9 & 4.1 & 0.3 & 0.27\\
L30003 & 500 & -12 & 0.5 & 45.5 & 2.0 & 1.3
\enddata
\tablecomments{$\mdot$ in units of $\msunyr$.
$R_{wall}$ is the radius of the wall at the inner edge of the dust disk,
corresponding to the dust sublimation radius for all except L316 and L30003.
$H_{wall}$ is the height of the wall that provides the best fit
to the observed SEDs; uncertainties in the exact level of the underlying
photosphere at $\lambda \sim 2-8$ \micron yield typical uncertainties
on $H_{wall}$ of about 0.1 $R_*$ (0.5-1 $R_*$ for the large holes). 
$H_{wall}^{pred}$ is the predicted wall height
for $a_{max} = 0.25$ \micron according to the treatment of
Muzerolle et al. (2004).}
\end{deluxetable}

\newpage

\begin{figure}
%\plotone{f1.eps}
\caption{Central $\sim$ 30'x30' portion of the 24 \micron map of IC 348.
The positions of all known members with spectral types M6 or later
are marked with green diamonds; those with measurable 24 \micron fluxes
are shown with yellow boxes.  (See PNG file.)
\label{bdpos}}
\end{figure}

%\begin{figure}
%\plotone{bdsed_mod_1mm.eps}
%\caption{Same as in Figure~\ref{sedism}, but with $a_{max}=1$ mm.
%\label{sed1mm}}
%\end{figure}

\end{document}